\documentclass[a4paper]{amsart}

\RequirePackage{amsmath}
\RequirePackage{bm}
\RequirePackage{amssymb}
\RequirePackage{upref}
\RequirePackage{amsthm}
\RequirePackage{enumerate}
\RequirePackage{pb-diagram}
\RequirePackage{amsfonts}
\RequirePackage[mathscr]{eucal}
\RequirePackage{verbatim}
\RequirePackage{xr}
\RequirePackage{graphicx}
\usepackage{calc}
\usepackage{xspace}
\RequirePackage{color}

\newcommand{\cf}{cf.\@\xspace}
\newcommand{\resp}{resp.\@\xspace}

\newcommand{\al}{\alpha}

\newcommand{\de}{\delta }

\newcommand{\h}{\eta}

\newcommand{\ka}{\kappa}
\newcommand{\lam}{\lambda}
\newcommand{\m}{\mu}

\newcommand{\Lam}{\varLambda}

\newcommand{\msp[1]}[1]{\mspace{#1mu}}

\newcommand{\R}[1][n+1]{{\protect\mathbb R}^{#1}}

\newcommand{\Cc}{{\protect\mathbb C}}

\newcommand{\N}{{\protect\mathbb N}}

\newcommand{\eR}{\stackrel{\lower1ex \hbox{\rule{6.5pt}{0.5pt}}}{\msp[3]\R[]}}
\newcommand{\eN}{\stackrel{\lower1ex \hbox{\rule{6.5pt}{0.5pt}}}{\msp[1]\N}}
\newcommand{\eO}{\stackrel{\lower1ex \hbox{\rule{6pt}{0.5pt}}}{\msc O}}

\newcommand\im{\implies}
\newcommand\ra{\rightarrow}

\newcommand{\un}{\infty}
\newcommand{\A}{\forall}

\newcommand{\set}[2]{\{\,#1\colon #2\,\}}
\newcommand{\uu}{\cup}

\newcommand{\uuu}{\bigcup}

\newcommand{\uud}{ \stackrel{\lower 1ex \hbox {.}}{\uu}}
\newcommand{\uuud}[1]{ \stackrel{\lower 1ex \hbox {.}}{\uuu_{#1}}}
\newcommand\su{\subset}

\newcommand\eS{\emptyset}
\newcommand{\sminus}[1][28]{\raise 0.#1ex\hbox{$\scriptstyle\setminus$}}

\newcommand{\wt}{\widetilde}

\newcommand{\wed}{\wedge}

\newcommand{\abs}[1]{\lvert#1\rvert}

\newcommand{\norm}[1]{\lVert#1\rVert}

\newcommand{\spd}[2]{\protect\langle #1,#2\protect\rangle}

\newcommand{\tit}{\textit}

\newcommand{\tup}{\textup}

\newcommand{\mc}{\protect\mathcal}
\newcommand{\msc}{\protect\mathscr}

\providecommand{\bysame}{\makebox[3em]{\hrulefill}\thinspace}

\newcommand{\bt}{\begin{thm}}
\newcommand{\bl}{\begin{lem}}
\newcommand{\bc}{\begin{cor}}
\newcommand{\bd}{\begin{definition}}
\newcommand{\bpp}{\begin{prop}}
\newcommand{\br}{\begin{rem}}
\newcommand{\bn}{\begin{note}}
\newcommand{\be}{\begin{ex}}
\newcommand{\bes}{\begin{exs}}
\newcommand{\bb}{\begin{example}}
\newcommand{\bbs}{\begin{examples}}
\newcommand{\ba}{\begin{axiom}}
\newcommand{\bas}{\begin{assumption}}

\newcommand{\et}{\end{thm}}
\newcommand{\el}{\end{lem}}
\newcommand{\ec}{\end{cor}}
\newcommand{\ed}{\end{definition}}
\newcommand{\epp}{\end{prop}}
\newcommand{\er}{\end{rem}}
\newcommand{\en}{\end{note}}
\newcommand{\ee}{\end{ex}}
\newcommand{\ees}{\end{exs}}
\newcommand{\eb}{\end{example}}
\newcommand{\ebs}{\end{examples}}
\newcommand{\ea}{\end{axiom}}
\newcommand{\eas}{\end{assumption}}

\newcommand{\bp}{\begin{proof}}
\newcommand{\ep}{\end{proof}}
\newcommand{\eps}{\renewcommand{\qed}{}\end{proof}}

\newcommand{\bal}{\begin{align}}

\newcommand{\bi}[1][1.]{\begin{enumerate}[\upshape #1]}
\newcommand{\bia}[1][(1)]{\begin{enumerate}[\upshape #1]}
\newcommand{\bin}[1][1]{\begin{enumerate}[\upshape\bfseries #1]}
\newcommand{\bir}[1][(i)]{\begin{enumerate}[\upshape #1]}
\newcommand{\bic}[1][(i)]{\begin{enumerate}[\upshape\hspace{2\cma}#1]}
\newcommand{\bis}[2][1.]{\begin{enumerate}[\upshape\hspace{#2\parindent}#1]}
\newcommand{\ei}{\end{enumerate}}

\newcommand\ndots{\raise 0.47ex \hbox {,}\hskip0.06em\cdots %
     \raise 0.47ex \hbox {,}\hskip0.06em} 

\newcommand{\q}{\quad}
\newcommand{\qq}{\qquad}

\newcommand\nd{\noindent}

\newskip\Csmallskipamount                                                
\Csmallskipamount=\smallskipamount
\newskip\Cmedskipamount
\Cmedskipamount=\medskipamount
\newskip\Cbigskipamount
\Cbigskipamount=\bigskipamount

\newcommand\cvs{\vspace\Csmallskipamount}   
\newcommand\cvm{\vspace\Cmedskipamount}

\newskip\csa
\csa=\smallskipamount

\newskip\cma
\cma=\medskipamount

\newskip\cba
\cba=\bigskipamount

\newdimen\spt
\newcommand\citem{\cvs\advance\itemno by
1{(\romannumeral\the\itemno})\hskip3pt}
\newcommand{\bitem}{\cvm\nd\advance\itemno by
1{\bf\the\itemno}\hspace{\cma}}

\newcount\itemno
\newcommand{\lae}[1]{\label{E:#1}}
\newcommand{\lat}[1]{\label{T:#1}}
\newcommand{\lal}[1]{\label{L:#1}}

\newcommand{\rt}[1]{Theorem~\ref{T:#1}}

\newcommand{\re}[1]{\eqref{E:#1}}
\newcommand{\frt}[1]{Theorem~\ref{T:#1} on page~\tup{\pageref{T:#1}}}

\newcommand{\fre}[1]{\eqref{E:#1} on page~\tup{\pageref{E:#1}}}

\newskip\thmskip
\thmskip=\parindent

\newskip\hsk
\newenvironment{hinw}{\labelsep=0pt\begin{list}{}{\labelsep=0pt\itemindent=0pt\labelwidth=0pt\leftmargin=\parindent\rightmargin=0pt\partopsep=\cba}%
\item\it\nopagebreak\nopagebreak}%
{\end{list}}

\newcommand\bh{\begin{hinw}}
\newcommand{\eh}{\end{hinw}}

\newtheoremstyle{normal}
  {\cba}
  {\cba}
  {}
  {\thmskip}
  {\bfseries}
  {.}
  {\hsk}
  {}

\newtheoremstyle{abschnitt}
  {\cba}
  {\cba}
  {}
  {\thmskip}
  {\bfseries}
  {.}
  {\hsk}
  {}

\newtheoremstyle{italic}
  {\cba}
  {\cba}
  {\itshape}
  {\thmskip}
  {\bfseries}
  {.}
  {\hsk}
  {}

\newtheoremstyle{aufgaben}
  {\cba}
  {\cba}
  {}
  {}
  {\normalsize\bfseries}
  {.}
  {\hsk}
  {}

\newtheoremstyle{break}
  {\cba}
  {\cba}
  {\itshape}
  {}
  {\bfseries}
  {.}
  {\newline}
  {}

\swapnumbers
\theoremstyle{italic}
\newtheorem{thm}[subsection]{Theorem}
\newtheorem{lem}[subsection]{Lemma}
\newtheorem{prop}[subsection]{Proposition}
\newtheorem{cor}[subsection]{Corollary}

\theoremstyle{normal}
\newtheorem{rem}[subsection]{Remark}
\newtheorem{definition}[subsection]{Definition}
\newtheorem{example}[subsection]{Example}
\newtheorem{examples}[subsection]{Examples}
\newtheorem{ex}[subsection]{Exercise}
\newtheorem{note}[subsection]{}
\newtheorem{axiom}[subsection]{Axiom}
\newtheorem{assumption}[subsection]{Assumption}

\theoremstyle{aufgaben}
\newtheorem{exs}[subsection]{Exercises}

\swapnumbers

\numberwithin{equation}{section}
\numberwithin{figure}{section}

\newenvironment{textequation}[1][0.8]
{\begin{equation}
\begin{aligned}
\begin{minipage}{#1\linewidth}}
{\end{minipage}
\end{aligned}
\end{equation}
\ignorespacesafterend}

\newcommand{\btext}{\begin{textequation}}
\newcommand{\etext}{\end{textequation}}

\def\hinweis{\@startsection{subsection}{2}%
 \z@{0.7\linespacing\@plus 0.5\linespacing}{0.7\linespacing}%
{\normalfont\itshape\indent}}

\usepackage[german,english]{babel}
\usepackage{graphicx}
\RequirePackage{amsmath}
\RequirePackage{bm}
\RequirePackage{amssymb}
\RequirePackage{upref}
\RequirePackage{amsthm}
\RequirePackage{enumerate}
\RequirePackage{pb-diagram}
\RequirePackage{amsfonts}
\RequirePackage[mathscr]{eucal}
\RequirePackage{verbatim}
\RequirePackage{xr}
\RequirePackage{graphicx}
\usepackage{calc}
\usepackage{xspace}

\makeatletter
\RequirePackage{color}
\newcommand{\ann}[1]{\renewcommand{\@makefnmark}{\mbox{$^{\color{red}{\@thefnmark}}$}}%
\footnote {#1}}
\makeatother

\RequirePackage{upref}
\RequirePackage{amsthm}
\RequirePackage{enumerate}
\usepackage[mathscr]{eucal}

\usepackage{xr-hyper}

\usepackage{calc}

\newlength{\oddsidemarginlength}
\newlength{\topmarginlength}

\newcounter{numberoflines}
\newcounter{tempcc}
\oddsidemargin=\oddsidemarginlength
\evensidemargin=\oddsidemargin
\topmargin=\topmarginlength
\begin{document}

\flushbottom


\title[Quantum cosmological Friedman models]{Quantum cosmological Friedman models with a Yang-Mills field and positive energy levels}

\author{Claus Gerhardt}
\address{Ruprecht-Karls-Universit\"at, Institut f\"ur Angewandte Mathematik,
Im Neuenheimer Feld 294, 69120 Heidelberg, Germany}
\email{gerhardt@math.uni-heidelberg.de}
\urladdr{http://www.math.uni-heidelberg.de/studinfo/gerhardt/}
\thanks{This work has been supported by the DFG}

%
\subjclass[2000]{35J60, 53C21, 53C44, 53C50, 58J05, 83C45}
\keywords{Quantum cosmology, Friedman model, big bang, Lorentzian manifold, Yang-Mills fields, general relativity, positive energy}
\date{\today}
%


\begin{abstract}
We prove the existence of a spectral resolution of the Wheeler-DeWitt equation when the matter field is provided by a Yang-Mills field, with or without mass term, if the spatial geometry of the underlying spacetime is homothetic to $\R[3]$. The energy levels of the resulting quantum model, i.e., the eigenvalues of the corresponding self-adjoint Hamiltonian with a pure point spectrum, are strictly positive.
\end{abstract}

\maketitle

\tableofcontents

\setcounter{section}{0}
\section{Introduction}

In a recent paper \cite{cg:qfriedman-ym} we proved the existence of a spectral resolution of the Wheeler-DeWitt equation when the matter field is provided by a massive Yang-Mills field. The underlying spacetimes could be either spatially \tit{closed}, i.e., spatially homothetic to $S^3$, or \tit{unbounded}, i.e., spatially homothetic to $\R[3]$.

However, the resulting quantum models had energy levels ranging from $-\un$ to $\un$, due to the employed techniques.

In the present paper we prove, in case that the underlying spacetime is spatially homothetic to $\R[3]$, a different spectral resolution the energy levels of which are strictly positive.

As we have explained in \cite[Introduction]{cg:qfriedman-ym} solving the Wheeler-DeWitt equation comprises three steps: First, the Hamilton operators corresponding to the gravitational field and the matter field, respectively, have to be separated; second, for one of the operators a complete set of eigenfunctions has to be found, i.e., a \tit{free} spectral resolution has to be proved without any constraints; third, for the remaining Hamilton operator then a \tit{constrained} spectral resolution has to be found by looking at the Wheeler-DeWitt equation as an \tit{implicit} eigenvalue problem.

In our previous paper we treated the open and closed spatial geometries simultaneously  and, therefore, had to use the Hamilton operator corresponding to the gravitational field to solve the free eigenvalue problem and the Hamilton operator corresponding to the Yang-Mills field to solve the implicit eigenvalue problem. For this reason we also had to assume a massive Yang-Mills field, since the scalar factor representing the mass played the role of the implicit eigenvalue. 

However, when assuming flat spatial sections a different approach is possible with positive energy levels. The technical difference is that now the gravitational Hamiltonian $H_1$ can be used to solve the implicit eigenvalue problem instead of the Hamiltonian of the Yang-Mills field.

The Wheeler-DeWitt equation had the form, \cf \cite[Theorem 3.2]{cg:qfriedman-ym},
\begin{equation}\lae{1.1}
H_2\psi-H_1\psi=0,
\end{equation}
where the wave function $\psi=\psi(r,y)$ belongs to a suitable subspace of $L^2(\R[]_+\times \R[],\Cc)$ and where
\begin{equation}\lae{1.2}
H_1\psi=-\Ddot\psi-\bar\Lam r^4\psi +4\tilde\ka r^2\psi,
\end{equation}
\begin{equation}\lae{1.3}
H_2\psi=-c_1\psi''+ V\psi -\bar\mu y^2 \psi.
\end{equation}

Here, the variables $r$ \resp $y$ represent the scale factor \resp the Yang-Mills field, and a dot indicates differentiation with respect to $r$ and a prime with respect to $y$. $\bar\Lam$ is a positive multiple of the cosmological constant $\Lam$, $\tilde\ka$ the spatial curvature, i.e., $\tilde\ka\in \{0,1\}$, $c_1$ a positive constant, $\bar\mu$ a positive multiple of the mass of the Yang-Mills field and $V$ the potential
\begin{equation}
V=2\al_M(\tilde\ka y+y^2)^2,
\end{equation}
where $\al_M$ is a positive coupling constant for the matter Lagrangian, \cf \cite[equ.\ (1.13)]{cg:qfriedman-ym}.

$H_1$ is the Hamiltonian of the gravitational field and $H_2$ the Hamiltonian of the (massive) Yang-Mills field.

Contrary to the situation in \cite{cg:qfriedman-ym} $\bar\mu$ will now be fixed, only subject to the requirement
\begin{equation}\lae{1.5}
\bar\mu<\bar\mu_0,
\end{equation}
where $0<\bar\mu_0$ is an extremal value such that the free eigenvalue problem
\begin{equation}
H_2\h=\mu\h
\end{equation}
will have a \tit{smallest} eigenvalue $\m=\mu_0=0$ when $\bar\mu=\bar\mu_0$ and $\tilde\ka=0$.

In case $\bar\mu<\bar\mu_0$ and $\tilde\ka=0$ the smallest eigenvalue $\mu_0$ will always be positive. We emphasize that especially the value $\bar\mu=0$ is allowed which would remove the mass term in the Lagrangian.

Choosing $\tilde\ka=0$ the Hamilton operator $H_1$ in \re{1.2} has the form
\begin{equation}
H_1 u=-\Ddot u-\bar\Lam r^4u
\end{equation}
and for this operator we can solve an implicit eigenvalue problem by using a rescaling trick as in \cite[Theorem 1.7]{cg:qfriedman}.

We shall prove:
\bt
Assuming $\tilde\ka=0$ and $\bar\mu$ satisfying \re{1.5}, there exists a self-adjoint operator $H$ in the Hilbert space  $L^2(\R[]_+\times\R[],\Cc)$,
\begin{equation}
H= H_2^{-1}\wt H_1=\wt H_1 H_2^{-1},
\end{equation}
where
\begin{equation}
\wt H_1 \psi=-\Ddot\psi+r^4\psi,
\end{equation}
 with a pure point spectrum consisting of countably many eigenvalues $\lam_{ij}$,
\begin{equation}
\lam_{ij}=\tilde\lam_i\mu_j^{-1},
\end{equation}
$\tilde\lam_i$ \resp $\mu_j$ are the eigenvalues of the operators $\tilde H_1$ \resp $H_2$,
 such that the properly rescaled eigenfunctions
 \begin{equation}
\psi_{ij}(r,y)=\wt\psi_{ij}(\lam_{ij}^{-\frac12}r,y)
\end{equation}
are solutions of the Wheeler-DeWitt equation
\begin{equation}
H_2\psi_{ij}-H_1\psi_{ij}=0,
\end{equation}
where
\begin{equation}
H_1\psi_{ij}=-\Ddot \psi_{ij}-\bar\Lam_{ij}r^4\psi_{ij}
\end{equation}
and
\begin{equation}
\bar\Lam_{ij}=-\lam_{ij}^{-3}.
\end{equation}
The eigenvalues $\lam_{ij}$ are strictly monotone increasing in $i$ and strictly monotone decreasing in $j$ and they range from $0$ to $\un$
\begin{equation}
\lim_i\lam_{ij}=\un\q\wed\q \lim_j\lam_{ij}=0.
\end{equation}
The solutions of the corresponding Schr\"odinger equation, with initial values $\tilde\psi_0$ belonging to the span of the eigenfunctions, provide a dynamical development of the quantum model.
\et

The  theorem will be proved in the following two sections.
\section{The eigenvalue problems}

The Hamiltonian in the Wheeler-DeWitt equation \fre{1.1} is already separated, hence, a separation of variables is possible
\begin{equation}
\psi(r,y)=u(r)\h(y),\qq (r,y)\in\R[]_+\times\R[].
\end{equation}

We first solve the free eigenvalue problem for $H_2$
\begin{equation}\lae{2.2}
H_2\h=\mu\h,
\end{equation}
where, for simplicity, we assume, without loss of generality, $\h$ to be real valued.

A complete set of eigenfunctions can be found with the help of a well-known variational principle, see e.g., \cite{cg:eigenwert} for details. Let $\mc H_2$ be the Hilbert space obtained by the completion of $C^\un_c(\R[])$ with respect to the norm
\begin{equation}
\norm{\h}^2_2=\int_{\R[]}(\abs{\h'}^2+y^4\abs \h^2).
\end{equation}

Then the quadratic form
\begin{equation}\lae{2.4}
K(\h)=\int_{\R[]}\abs\h^2
\end{equation}
is compact in $\mc H_2$, \cf \cite[Lemma 6.8]{cg:qfriedman} for a proof in a similar situation, and the quadratic form
\begin{equation}
\spd{H_2\h}\h+c K(\h)=\int_{\R[]}(c_1\abs{\h'}^2+V\abs \h^2-\bar\mu y^2\abs\h^2)+cK(\h)
\end{equation}
is uniformly positive definite if the positive constant $c$ is large enough.

Thus, we conclude:
\bt\lat{2.1}
There exist countably many eigenfunctions $\h_i$ with eigenvalues $\mu_i$ such that
\begin{equation}
\mu_i<\mu_{i+1}\qq\A\,i\in\N
\end{equation}
and
\begin{equation}
\lim_i\mu_i=\un.
\end{equation}
The eigenfunctions $(u_i)$ are dense in $\mc H_2$ as well as in $L^2(\R[])$ and the eigenvalues have multiplicities $1$.
\et

The theorem is valid for arbitrary values of $\bar\mu$ in \fre{1.3}. Moreover, we can prove a very precise \tit{mass gap} for that particular Yang-Mills Hamiltonian:
\bt
There exists exactly one $\bar\mu_0>0$ such that, when choosing $\bar\mu=\bar\mu_0$ in \fre{1.3}, the corresponding smallest eigenvalue $\mu_0$ satisfies
\begin{equation}
\mu_0=0.
\end{equation}
Choosing $\bar\mu<\bar\mu_0$ the corresponding smallest eigenvalue $\mu_0$ is strictly positive
\begin{equation}
\mu_0=\mu(\bar\mu)>0.
\end{equation}
\et

\bp
(i) For $\bar\mu\in\R[]$ and $\h\in \mc H_2$ consider the functional
\begin{equation}
J_{\bar\mu}(\h)=\int_{\R[]}c_1\abs{\h'}^2+V\abs\h^2-\bar\mu\int_{\R[]}y^2\abs\h^2.
\end{equation}
Define
\begin{equation}
\mu(\bar\mu)=\inf\set{J_{\bar\mu}(\h)}{\int_{\R[]}\abs\h^2=1,\;\h\in\mc H_2}
\end{equation}
and set
\begin{equation}
E=\set{\bar\mu}{\mu(\bar\mu)\le 0}.
\end{equation}

We immediately deduce
\begin{equation}
E\ne \eS
\end{equation}
and
\begin{equation}\lae{2.14}
\bar\mu\in E\im \bar\mu>0.
\end{equation}

We also note that $\mu(\bar\mu)$ is exactly the smallest eigenvalue $\mu_0$ of the corresponding eigenvalue problem \re{2.2}
\begin{equation}
\mu_0=\mu(\bar\mu).
\end{equation}

Let
\begin{equation}
\bar\mu_0=\inf E,
\end{equation}
then $\bar\mu_0\in E$ because of the compactness of the form \re{2.4} and hence
\begin{equation}
\bar\mu_0>0,
\end{equation}
in view of \re{2.14}.

\cvm
(ii) Next, we claim that
\begin{equation}
\mu_0=\mu(\bar\mu_0)=0.
\end{equation}

We argue by contradiction. Assume
\begin{equation}
\mu_0<0
\end{equation}
and let $\h$ be a corresponding eigenfunction with unit $L^2$-norm such that
\begin{equation}
J_{\bar\mu_0}(\h)=\mu_0<0,
\end{equation}
then we infer
\begin{equation}
J_{\bar\mu_0-\de}(\h)\le\tfrac{\mu_0}2<0,
\end{equation}
if $\de>0$ is small enough contradicting the definition of $\bar\mu_0$.

\cvm
(iii) Let $\bar\mu<\bar\mu_0$, then for any $0\ne\h\in\mc H_2$
\begin{equation}
0\le J_{\bar\mu_0}(\h)<J_{\bar\mu}(\h),
\end{equation}
hence
\begin{equation}
\mu(\bar\mu)>0.
\end{equation}

(iv) To prove the uniqueness of $\bar\mu_0$ let $\bar\mu_1\ne\bar\mu_0$ be another value such that
\begin{equation}
\mu(\bar\mu_1)=0.
\end{equation}
In view of (iii) there holds
\begin{equation}
\bar\mu_0<\bar\mu_1,
\end{equation}
hence
\begin{equation}
0\le J_{\bar\mu_1}(\h)<J_{\bar\mu_0}(\h)\qq\A\,0\ne\h\in\mc H_2;
\end{equation}
a contradiction.
\ep

Next, we consider the constrained eigenvalue problem for $H_1$. Let $\mu>0$ be one of the eigenvalues of $H_2$, then we look at the implicit eigenvalue problem
\begin{equation}\lae{2.27}
H_1u=-\Ddot u-\bar\Lam r^4u=\mu u,
\end{equation}
where $\bar\Lam$ or  $-\bar\Lam$ should play the role of an eigenvalue, i.e., it is more precisely an implicit eigenvalue problem for the operator
\begin{equation}
u\ra -\Ddot u -\mu u.
\end{equation}
However, the quadratic form
\begin{equation}
K(u)=\int_{\R[]_+}r^4\abs u^2
\end{equation}
is not compact relative to any reasonable energy form.

To solve \re{2.27} we have to use a rescaling trick as in \cite[Theorem 1.7]{cg:qfriedman}.

Let us first consider the Hamiltonian
\begin{equation}
\widetilde H_1u=-\Ddot u+r^4 u
\end{equation}
with corresponding energy form
\begin{equation}
\spd{\wt H_1u}u=\int_{\R[]_+}(\abs{\dot u}^2+r^4\abs u^2)\equiv \norm u_1^2
\end{equation}
and define the real Hilbert space $\mc H_1$ as the completion of $C^\un_c(\R[]_+)$ with respect to the norm $\norm\cdot_1$.

The eigenvalue problem
\begin{equation}
\wt H_1 \tilde u=\tilde \lam\tilde u
\end{equation}
is then solvable and we obtain an analogue of \rt{2.1}, namely:
\bt\lat{2.3}
There exist countably many eigenfunctions $\tilde u_i$ with eigenvalues $\tilde\lam_i$ such that
\begin{equation}
\tilde\lam_i<\tilde\lam_{i+1}\qq\A\,i\in\N,
\end{equation}
\begin{equation}
\tilde\lam_0>0,
\end{equation}
and
\begin{equation}
\lim_i\tilde\lam_i=\un.
\end{equation}
The eigenfunctions $(\tilde u_i)$ are dense in $\mc H_1$ as well as in $L^2(\R[]_+)$ and the eigenvalues have multiplicities $1$. 
\et

\bt\lat{2.4}
Let $\mu>0$, then the pairs $(\tilde u_i,\lam_i)$ represent a complete set of eigenfunctions with eigenvalues
\begin{equation}
\lam_i=\tilde\lam_i\mu^{-1}
\end{equation}
for the eigenvalue problem
\begin{equation}
\wt H_1 u=\lam \mu u.
\end{equation}
The rescaled functions
\begin{equation}
u_i(r)=\tilde u_i(\lam_i^{-\frac12}r)
\end{equation}
then satisfy
\begin{equation}
-\Ddot u+\lam_i^{-3}r^4u_i=\mu u_i,
\end{equation}
or, if we set
\begin{equation}
\bar\Lam_i=-\lam_i^{-3},
\end{equation}
\begin{equation}
-\Ddot u-\bar\Lam_ir^4u_i=\mu u_i.
\end{equation}
\et
\section{The spectral resolution}

Let $(\mu,\h)$ \resp $(\lam,\tilde u)$ satisfy
\begin{equation}
H_2\h=\mu\h
\end{equation}
\resp
\begin{equation}
\wt H_1\tilde u=\lam \mu \tilde u,
\end{equation}
then
\begin{equation}
\tilde\psi=\tilde u\h
\end{equation}
solves
\begin{equation}\lae{3.4}
\wt H_1\tilde\psi=\lam H_2\tilde\psi,
\end{equation}
or equivalently, in view of \frt{2.4},
\begin{equation}
H_1\psi-H_2\psi=0,
\end{equation}
where
\begin{equation}
\psi=u\h,
\end{equation}
\begin{equation}
u(r)=\tilde u(\lam^{-\frac12}r),
\end{equation}
\begin{equation}
H_1\psi=-\Ddot\psi-\bar\Lam r^4\psi,
\end{equation}
and
\begin{equation}
\bar\Lam =- \lam^{-3},
\end{equation}
i.e., $\psi$ is a solution of the Wheeler-DeWitt equation.

Moreover,
\begin{equation}\lae{6.5}
\dot\psi=\dot u\h\q\wed\q \psi'=u\h',
\end{equation}
hence,
\begin{equation}\lae{6.6}
\int_{\R[]_+\times\R[]}\abs{D\psi}^2=\int_{\R[]_+}\abs{\dot u}^2\int_{\R[]}\abs\h^2 + \int_{\R[]_+}\abs u^2\int_{\R[]}\abs{\h'}^2,
\end{equation}
and similarly,
\begin{equation}\lae{6.7}
\int_{\R[]_+\times\R[]}\abs\psi^2y^p=\int_{\R[]_+}\abs u^2\int_{\R[]}\abs\h^2y^p,
\end{equation}
for $p=2,4$, as well as
\begin{equation}
\int_{\R[]_+\times\R[]}\abs\psi^2r^4=\int_{\R[]_+}\abs u^2r^4\int_{\R[]}\abs\h^2.
\end{equation}

Thus, $\psi$ has bounded norm
\begin{equation}
\norm\psi^2=\int_{\R[]_+\times\R[]}\abs{D\psi}^2+\int_{\R[]_+\times\R[]}\abs\psi^2(r^4+y^4).
\end{equation}

Let $\mc H$ be the completion of $C^\un_c(\R[]_+\times\R[])$ with respect to this norm, then $\mc H$ can be viewed as  a dense subspace of  
\begin{equation}
\mc H_0=L^2(\R[]_+\times\R[])
\end{equation}
and the eigenfunctions  of \re{3.4} are complete in $\mc H$ as well as $\mc H_0$, where we note that the eigenfunctions $\tilde\psi_{ij}$ are products
\begin{equation}
\tilde\psi_{ij}=\tilde u_i\h_j
\end{equation}
with eigenvalues
\begin{equation}
\lam_{ij}=\tilde\lam_i\mu_j^{-1},
\end{equation}
where we recall that $\tilde\lam_i$ are the eigenvalues of the Hamiltonian $\wt H_1$, \cf \frt{2.4}. Thus, the eigenvalues $\lam_{ij}$ are strictly monotone increasing in $i$ and strictly monotone decreasing in $j$ and they range from $0$ to $\un$
\begin{equation}
\lim_i\lam_{ij}=\un\q\wed\q \lim_j\lam_{ij}=0.
\end{equation}

The claim that the eigenfunctions are complete needs some verification.
\bl\lal{5.10}
The eigenfunctions $\tilde\psi_{ij}$ are complete in $\mc H$ as well as in $\mc H_0$.
\el

\bp
It suffices to prove the density in $\mc H$. The eigenfunctions are certainly complete in the closure of $C^\un_c(\R[]_+)\otimes C^\un_c(\R[])$ in $\mc H$, in view of \re{6.5} and \re{6.6}, but $C^\un_c(\R[]_+)\otimes C^\un_c(\R[])$ is dense in $\mc H$ as can be easily proved with the help of the Weierstra{\ss} approximation theorem.  
\ep

From now on we shall assume that the functions are complex valued. Denote by $A$ the operator
\begin{equation}
A=H_2^{-1}\wt H_1
\end{equation}
with domain $D(A)\su\mc H_0$ equal to the subspace generated by its eigenfunctions $\tilde \psi_{ij}$.

We observe that $A$ is well defined and that
\begin{equation}
H_2^{-1}\wt H_1=\wt H_1 H_2^{-1}.
\end{equation}
Moreover, one easily checks that $H_2^{-1}$ and hence $A$ are symmetric.

\bl
$A$ is essentially self-adjoint in $\mc H_0$.
\el
\bp
It suffices to prove that $R(A\pm i)$ is dense, which is evidently the case, since the eigenfunctions belong to $R(A\pm i)$.
\ep

Let $H$ be the closure of $A$, then $H$ is self-adjoint and the spectral resolution for the Wheeler-DeWitt equation accomplished, since there holds:
\bl
Let $(\psi,\lam)\in \mc H\times \R[]_+$ be a solution of the Wheeler-DeWitt equation
\begin{equation}
-\Ddot\psi+\lam^{-3}r^4\psi-H_2\psi=0,
\end{equation}
then there exists $(ij)\in\N\times\N$ such that
\begin{equation}
\lam=\lam_{ij}
\end{equation}
and
\begin{equation}
\psi=\psi_{ij},
\end{equation}
where
\begin{equation}
\psi_{ij}(r,y)=\tilde\psi_{ij}(\lam_{ij}^{-\frac12}r,y),
\end{equation}
and $\tilde\psi_{ij}$ is an eigenfunction of $H$ with eigenvalue $\lam_{ij}$.
\el

\bp
Define
\begin{equation}
\tilde\psi (r,y)=\psi(\lam^{\frac12}r,y),
\end{equation}
then $\tilde\psi$ is a solution of
\begin{equation}
H\tilde\psi=\lam\tilde\psi,
\end{equation}
hence the result. 

Note that the eigenspaces of $H$ are not necessarily one-dimensional.
\ep

The Schr\"odinger equation for $H$ offers a dynamical development of the system provided the initial value is a finite superposition of eigenfunctions, since then the  time dependent solutions are also solutions of the Wheeler-DeWitt equation, \cf the remarks at the end of \cite[Section 8]{cg:qfriedman}.


\providecommand{\bysame}{\leavevmode\hbox to3em{\hrulefill}\thinspace}
\providecommand{\href}[2]{#2}



\end{document}